\documentclass[aps,prb,twocolumn,preprintnumbers,amsmath,amssymb,superscriptaddress]{revtex4}%

\usepackage{graphicx}
\usepackage{dcolumn}
\usepackage{amsmath}
\usepackage{color}
\usepackage{hyperref}

\newcommand{\RN}[1]{\textup{\uppercase\expandafter{\romannumeral#1}}}%

\begin{document}

\title{In-plane magnetic penetration depth in Sr$_2$RuO$_4$: muon-spin rotation/relaxation study}
\author{Rustem Khasanov}
 \email{rustem.khasanov@psi.ch}
 \affiliation{Laboratory for Muon Spin Spectroscopy, Paul Scherrer Institut, CH-5232 Villigen PSI, Switzerland}

\author{Aline Ramires}
 \affiliation{Laboratory for Theoretical and Computational Physics, Paul Scherrer Institut, CH-5232 Villigen PSI, Switzerland}

\author{Vadim Grinenko}
 \affiliation{Tsung-Dao Lee Institute, Shanghai Jiao Tong University, Pudong, Shanghai, China}

\author{Ilya Shipulin}
 \affiliation{Leibniz Institute for Solid State and Materials Research, D-01069, Dresden, Germany}

\author{Naoki Kikugawa}
 \affiliation{National Institute for Materials Science, Tsukuba, Ibaraki 305-0003, Japan}

\author{Dmitry A. Sokolov}
  \affiliation{Max Planck Institute for Chemical Physics of Solids, N\"{o}thnitzer Strasse 40, 01187 Dresden, Germany}

\author{Yoshiteru Maeno}
 \affiliation{Department of Physics, Kyoto University, Kyoto 606-8502, Japan}
 \affiliation{Toyota Riken - Kyoto University Research Center (TRiKUC), Kyoto 606-8501, Japan}

\author{Hubertus Luetkens}
 \affiliation{Laboratory for Muon Spin Spectroscopy, Paul Scherrer Institut, CH-5232 Villigen PSI, Switzerland}

\author{Zurab Guguchia}
 \affiliation{Laboratory for Muon Spin Spectroscopy, Paul Scherrer Institut, CH-5232 Villigen PSI, Switzerland}

\begin{abstract}
We report on measurements of the in-plane magnetic penetration depth ($\lambda_{\rm ab}$) in single crystals of Sr$_2$RuO$_4$ down to $\simeq 0.015$~K by means of muon-spin rotation/relaxation. The linear temperature dependence of $\lambda^{-2}_{\rm ab}$ for $T\lesssim 0.7$~K suggests the presence of nodes in the superconducting gap. This statement is further substantiated by observation of the Volovik effect, {\it i.e.} the reduction of $\lambda_{ab}^{-2}$ as a function of the applied magnetic field.
The experimental zero-field and zero-temperature value of $\lambda_{\rm ab}=124(3)$~nm
agrees with $\lambda_{\rm ab}\simeq 130$~nm, calculated based on results of electronic structure measurements reported in [Phys. Rev X {\bf 9}, 021048 (2019)]. Our analysis reveals that a simple nodal superconducting energy gap, described by the lowest possible harmonic of a gap function, does not capture the dependence of $\lambda_{\rm ab}^{-2}$ on $T$, so the higher angular harmonics of the energy gap function need to be introduced.
\end{abstract}

%\pacs{74.70.Xa, 74.25.Bt, 74.45.+c, 76.75.+i}
\maketitle

%Introduction

The unconventional superconductivity in Sr$_2$RuO$_4$ attracts tremendous attention even after 30 years of its discovery.\cite{Maeno_Nature_1994} The pure compound superconducts below $\simeq 1.5$~K\cite{Mao_MRB_2000} and has a relatively simple, quasi two-dimensional electronic structure with three nearly cylindrical Fermi surface sheets.\cite{Mackenzie_PRL_1996, Bergemann_PRL_2000, Damascelli_PRL_2000, Burganov_PRL_2016, Tamai_PRX_2019, Morales_PRL_2023} In addition, high quality crystals of Sr$_2$RuO$_4$ can be produced in large volumes\cite{Mao_MRB_2000, Bobowski_CM_2019} and the samples are  clean,\cite{Mackenzie_PhysC_1996, Maeno_JPSJ_1997, Bergemann_AdPh_2003} meaning that disorder is a less complication  factor in experiments as it is for most other unconventional superconductors.

One particular question to be answered is the symmetry of the superconducting order parameter.  The linearity of the electronic specific heat capacity at low temperatures,\cite{Nizhaki_JPCM_2000}  the residual linear term of the in-plane thermal conductivity,\cite{Suzuki_PRL_2002} the attenuation rate of ultrasound,\cite{Lupien_JPSJ_2001} and the field-oriented specific heat measurements \cite{Deguchi_PRL_2004} suggest the existence of nodes (or deep minima) in the superconducting  gap. Recent thermal conductivity measurements for heat currents flowing parallel and perpendicular to the crystallographic $c-$axis further indicate that these nodes/minima are oriented parallel to the $c-$axis.\cite{Hassinger_PRX_2017} This finding was also supported by measurements of Bogolubov quasiparticle interference, which allowed an identification of the position of the gap nodes in two Fermi surface sheets, and an estimate of the maximum value of the superconducting gap $\Delta^{\rm max}\simeq 0.35$~meV.\cite{Sharma_PNAS_2020}

Given the temperature dependence of the specific heat and thermal conductivity experiments,\cite{Nizhaki_JPCM_2000, Suzuki_PRL_2002, Hassinger_PRX_2017} one expects a linear behavior of $\lambda^{-2}(T)$ at low temperatures. However, all experiments  existing up to date report a power law dependence $\lambda^{-2}(T)\propto 1-(T/T_{\rm c})^n$ with the exponent $n$ ranging between 2 and 3.\cite{Luke_PhysB_2000, Bonalde_PRL_2000, Aegerter_JPCM_1998, Baker_PRB_2009} Such behavior was suggested to be consistent with the presence of nodes in the superconducting gap and ascribed to nonlocal electrodynamic effects leading to $T^2$ instead of linear behavior of $\lambda^{-2}(T)$ at low temperatures.\cite{Bonalde_PRL_2000, Kusunose_EPL_2002, Kubo_JPSJ_2000}
In this work, we report on measurements of the in-plane magnetic penetration depth ($\lambda_{\rm ab}$) in high-quality single crystals of Sr$_2$RuO$_4$ ($T_{\rm c}\simeq 1.4$~K, Ref.~\onlinecite{Supplememntal_part}) down to $T\simeq 15$~mK by means of muon-spin rotation/relaxation ($\mu$SR).
The observed linear temperature dependence of $\lambda_{\rm ab}^{-2}$ for $T\lesssim T_{\rm c}/2$ is consistent with the presence of nodes in the superconducting energy gap.

%Experiment and discussions

\begin{figure*}[htb]
\includegraphics[width=0.95\linewidth]{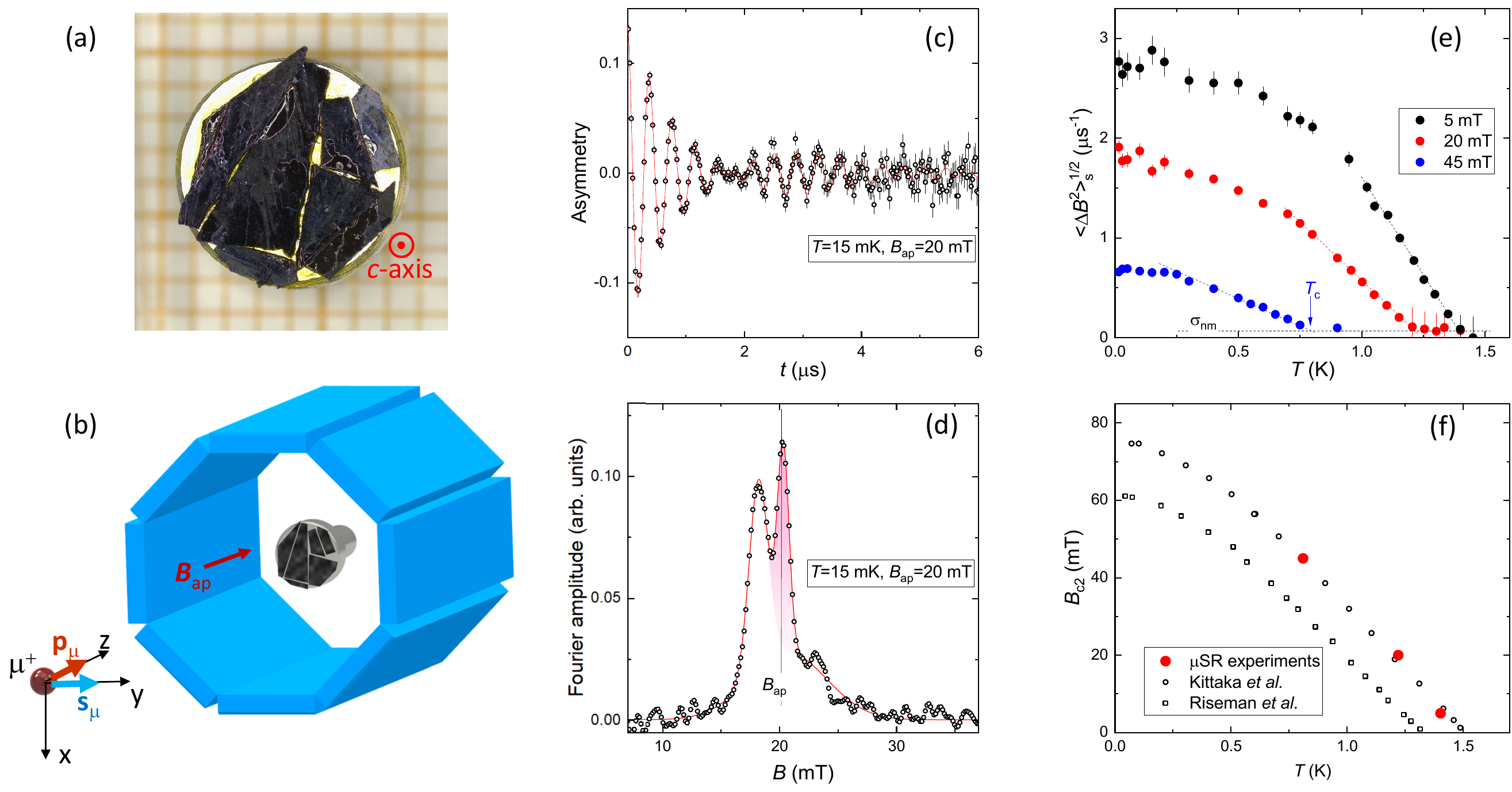}
\caption{ (a) The $c-$axis oriented Sr$_2$RuO$_4$ single crystals mounted on the $\mu$SR sample holder. (b) Schematic view of the experimental setup. Eight positron detectors are arranged in a ring surrounding the sample. The external magnetic field is applied along the muon momentum and parallel to crystal's $c-$axis (${\bf B}_{\rm ap} \parallel {\bf p}_\mu \parallel c$).  The muon-spin ${\bf s }_\mu$ stays perpendicular to the applied magnetic field (${\bf B}_{\rm ap} \perp {\bf s}_\mu$). (c) Muon-time spectra collected at $T=15$~mK and $B_{\rm ap}=20$~mT. (d) Fourier transform of the muon-time spectra shown in panel (c). The red area denotes the background contribution. The solid line in (c) and (d) are  three-component fits.\cite{Supplememntal_part} (e) Temperature dependence of the square root of the second moment $\langle {\Delta B^2}\rangle_{\rm s}^{1/2}$ measured at $B_{\rm ap}=5$, 20, and 45~mT. The transition temperature $T_{\rm c}$ at $B=B_{\rm ap}$ is defined by the intersection of the linearly extrapolated  $\langle \Delta B^2 \rangle_{\rm s}^{1/2}(T, B_{\rm ap})$ close to $T_{\rm c}$ with the nuclear moment contribution, $\sigma_{\rm nm}=const$ line. (f) Temperature dependence of the upper critical field $B_{\rm c2}^\parallel$ obtained by Kittaka {\it et al.}\cite{Kitaka_PRB_2009} and Riseman {\it et al.}\cite{Riseman_Nature_1998}. The red points are $B_{\rm c2}^\parallel(T)$ values from this work.}
 \label{fig:experiment}
\end{figure*}

Single crystals of Sr$_2$RuO$_4$ were grown by the floating zone technique.\cite{Bobowski_CM_2019, Kikugawa_Crystals_2021} Several crystals were cleaved from  rod n30 and mounted on the $\mu$SR sample holder [Fig.~\ref{fig:experiment}~(a)]. $\mu$SR experiments were carried out at the $\pi$E3 beamline using the HAL-9500 spectrometer (Paul Scherrer Institute, Switzerland). A schematic representation of the experimental setup is shown in Fig.~\ref{fig:experiment}~(b).
The experimental data were analyzed using the MUSRFIT package.\cite{MUSRFIT}

Figure \ref{fig:experiment}~(c) shows the transverse-field (TF) $\mu$SR time spectra collected at $T=15$~mK under the applied field $B_{\rm ap}=20$~mT. The corresponding Fourier transform, reflecting the internal field distribution [$P(B)$] in Sr$_2$RuO$_4$ in the superconducting state, is presented in Fig.~\ref{fig:experiment}~(d). The sharp peak at $B=B_{\rm ap}$ represents the residual background signal from muons missing the sample. The asymmetric $P(B)$ distributions possess the basic features expected for an ordered flux line lattice (FLL), namely: the cutoff at low fields, the peak shifted below $B_{\rm ap}$ and the tail towards the high field direction.\cite{Maisuradze_JPCM_2009, Khasanov_PRB_2016}
The $\mu$SR time spectra  were fitted by using a three-component Gaussian expression  with the first (the temperature and field independent) component corresponding to the background contribution $P_{\rm b}(B)$ and another two components accounting for the asymmetric distribution $P_{\rm s}(B)$ within the sample: $P(B)=P_{\rm b}(B)+P_{\rm s}(B)$.\cite{Supplememntal_part, Khasanov_PRL_2007, Khasanov_PRB_2016}

The temperature dependence of a square root of the second central moment of $P_{\rm s}(B)$ measured at $B_{\rm ap}=5$, 20 and 45~mT [$\langle \Delta B^2 \rangle_{\rm s}^{1/2}(T, B_{\rm ap})$] is presented in Fig.~ \ref{fig:experiment}~(e).  The nuclear dipole field contribution $\sigma_\text{nm}$ is assumed to be temperature-independent, and determined from data at $T > T_{\rm c}$ [the horizontal solid line in Fig.~ \ref{fig:experiment}~(e)].  The superconducting component $\langle \Delta B^2 \rangle_{\rm sc}$ is then obtained by subtracting $\sigma_{\rm nm}$ from the measured second moment: $\langle \Delta B^2 \rangle_{\rm sc} =\langle \Delta B^2 \rangle_{\rm s} - \sigma_{nm}^2$.  Following Refs.~\onlinecite{Brandt_PRB_1988, Brandt_PRB_2003}, $\langle \Delta B^2 \rangle_{\rm sc}$ is a function of the magnetic penetration depth $\lambda$ and  the reduced field $b=B/B_{\rm c2}$:
\begin{eqnarray}
\langle \Delta B^2 \rangle_{\rm sc}^{1/2} [\mu{\rm s}^{-1}] & = & A\; (1-b) [1+1.21 \nonumber \\
& &\times(1-\sqrt{1-b})^3] \; \lambda^{-2} [\mu{\rm m}^{-2}].
 \label{eq:Brandt}
\end{eqnarray}
Here $B_{\rm c2}$ is the upper critical field and $A$ is a coefficient capturing the FLL symmetry. $A$ is $\simeq4.83$ and $5.07$ for the 'hexagonal` and 'square` FLL, respectively.\cite{Brandt_PRB_2003,Khasanov_MoSb_PRB_2008} Note that, since in our experiments the magnetic field was applied along the crystallographic $c-$axis [Figs.~\ref{fig:experiment}~(a) and (b)], $\langle \Delta B^2 \rangle_{\rm sc}^{1/2}$ is a function of the in-plane magnetic penetration depth $\lambda_{\rm ab}$ and $B_{\rm c2}$ in ${\bf B}_{\rm ap}\parallel c$ orientation ($B_{\rm c2}^\parallel$).
Following Eq.~\ref{eq:Brandt}, the temperature dependence of $\lambda_{\rm ab}^{-2}$ can be reconstructed from the measured $\langle \Delta B^2 \rangle_{\rm sc}^{1/2}(T,B_{\rm ap})$ if $B_{\rm c2}^\parallel(T)$ is known. Figure~\ref{fig:experiment}~(f) compares $B_{\rm c2}^\parallel$ obtained from $\langle \Delta B^2 \rangle_{\rm s}^{1/2}(T)$ curves measured at $B_{\rm ap}=5$, 20, and 45~mT with the literature data.\cite{Kitaka_PRB_2009, Riseman_Nature_1998} Good agreement is obtained with $B_{\rm c2}^\parallel(T)$ from Kittaka {\it et al.}\cite{Kitaka_PRB_2009}, which was further used in the $\lambda_{\rm ab}^{-2}(T)$ reconstruction procedure. The FLL was assumed to be of a 'square` symmetry, as reported in Refs.~\onlinecite{Riseman_Nature_1998, Aegerter_JPCM_1998, Ray_PRB_2014} for $B_{\rm ap}\gtrsim3.5$~mT.

\begin{figure}[htb]
\includegraphics[width=1.0\linewidth]{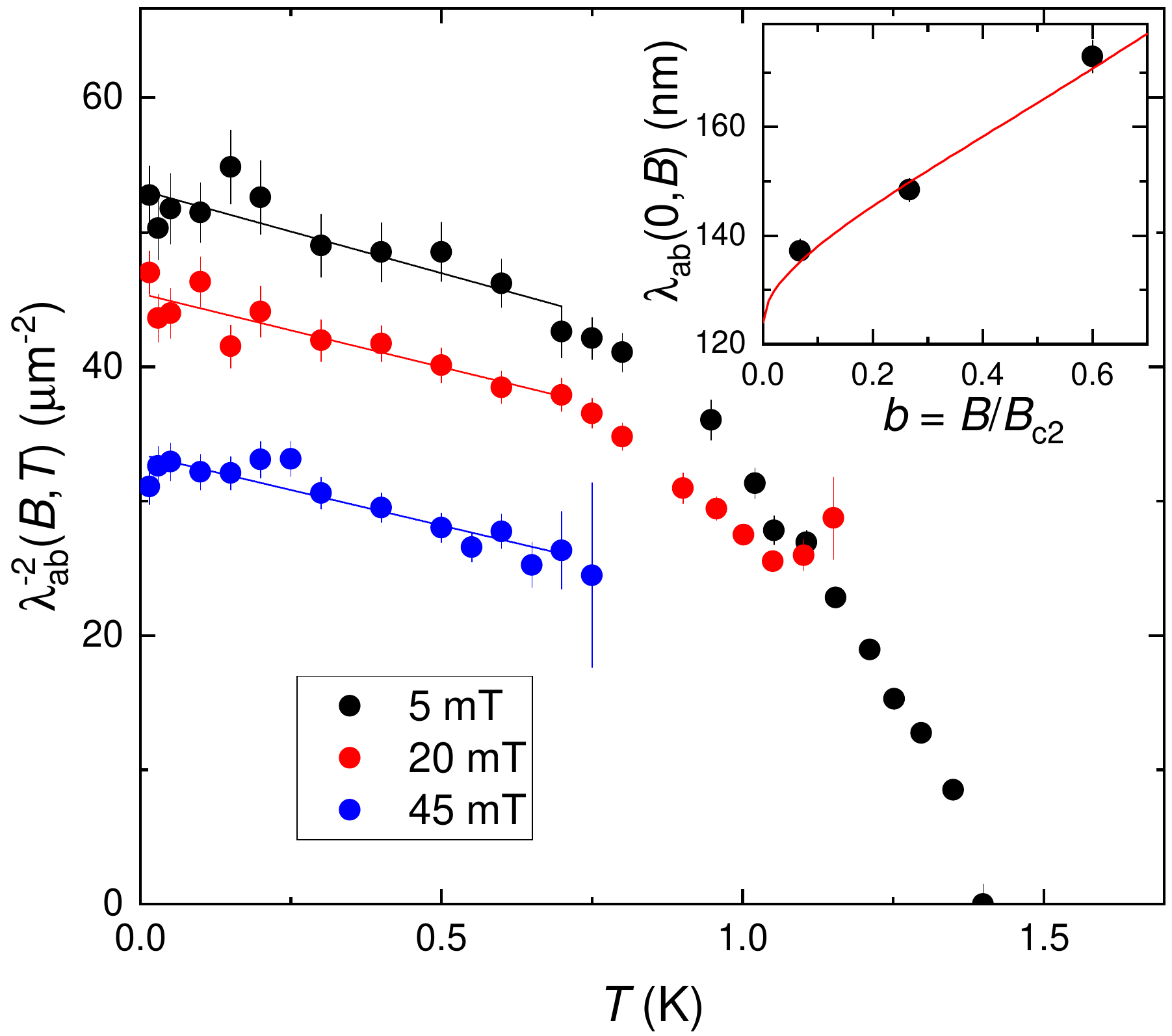}
\caption{$\lambda_{\rm ab}^{-2}$ {\it vs.} $T$ curves for Sr$_2$RuO$_4$ reconstructed from the measured $\langle \Delta B^2 \rangle_s^{1/2} (T)$'s presented in Fig.~\ref{fig:experiment}~(e). The solid lines are linear fits for $T\lesssim 0.7$~K. The inset shows the dependence of the zero-temperature $\lambda_{\rm ab}(0)$ on the reduced magnetic field $b=B/B_{\rm c2}$. }
 \label{fig:lambda_vs_T}
\end{figure}

Figure~\ref{fig:lambda_vs_T} shows $\lambda_{\rm ab}^{-2}$ {\it vs.} $T$  for $B_{\rm ap}=5$, 20, and 45~mT.  From these data, the following three important points emerge:
  (i) Below $T\simeq 0.7$~K ({\it i.e} for $T\lesssim T_{\rm c}/2$) $\lambda_{\rm ab}^{-2}(T)$ depends linearly on temperature. The linear fits result in similar values of ${\rm d}\lambda_{\rm ab}^{-2}/{\rm d} T$ slopes, which is a clear indication of the presence of nodes in the superconducting  gap;
  (ii) For $T\lesssim 0.7$~K the $\lambda_{\rm ab}^{-2}(T)$ slopes are nearly parallel, suggesting   $\lambda_{\rm ab}^{-2}(T,B_{\rm ap})$ values  to decrease with increasing $B_{\rm ap}$. This is a manifestation of the so-called Volovik effect, which is caused by magnetic field-induced quasiparticle excitations around gap nodes;\cite{Volovik_JETP-L_1993, Kopnin_JETP-L_1996}
  (iii) The absolute value of the zero-temperature and the zero-field $\lambda_{\rm ab}(0,0)$  could be obtained from its dependence   on the reduced field $b$ (see the inset in Fig.~\ref{fig:lambda_vs_T}).  According to Volovik,\cite{Volovik_JETP-L_1993} the density of excited states increases proportionally to $\sqrt{b}$, leading to:\cite{Kadono_JPCM_2004, Khasanov_PRB_2009}
      \begin{equation}
      \lambda(b)/\lambda(b=0)=(1-K\sqrt{b})^{1/2}.
       \label{eq:lambda_vs_b}
      \end{equation}
  Here $K$ captures the strength of the Volovik effect. The solid line in the inset of Fig.~\ref{fig:lambda_vs_T} corresponds to the fit of Eq.~\ref{eq:lambda_vs_b} to the data with $K=0.61(1)$ and  $\lambda_{\rm ab}(0, 0)=124(3)$~nm.
To summarise, the results presented in Fig.~\ref{fig:lambda_vs_T} suggest the presence of nodes in the energy gap structure of Sr$_2$RuO$_4$, in agreement  with the temperature dependencies of the thermal power and electronic specific heat, reported in Refs.~\onlinecite{Nizhaki_JPCM_2000, Suzuki_PRL_2002, Hassinger_PRX_2017}.

%Note that, at first glance, our data disagree with the results of Bonalde {\it et al.}\cite{Bonalde_PRL_2000} [{\it i.e.} the most complete $\lambda(T)$ measurements of Sr$_2$RuO$_4$ up to date], reporting a quadratic, rather than linear, low-temperature behavior of $\lambda^{-2}(T)$. This is, however, not the case. The fit of a $d-$wave gap model result in agreement with $\lambda^{-2}(T)$ data from Ref.~\onlinecite{Bonalde_PRL_2000} down to $T\lesssim 0.1\; T_{\rm c}=0.15$~K. This may suggest that the characteristic temperature $T^\ast$, which is associated with the change of $\rho_{\rm s}(T)$ from a nodal to $T^{2}$ like behaviour, might be shifted to lower values, {\it i.e.} down to $T^\ast\simeq0.15$~K, compared to $T^\ast\simeq0.82$~K from Ref.~\onlinecite{Bonalde_PRL_2000}.

The zero-temperature value of the in-plane penetration depth
can also be calculated from the electronic band dispersion. For a quasi-two-dimensional superconductor $\lambda^{-2}(0)$ can be written as:\cite{Evtushinsky_NPJ_2009}
\begin{equation}
\lambda_{\rm ab}^{-2}(0) \equiv \sum_i \lambda_{{\rm ab,}i}^{-2}(0) = \frac{e^2}{2\pi \varepsilon_0 c^2 h L_{\rm c}}  \sum_i  \oint v_{{\rm F},i}({\bf k})
{\rm d}k.
\label{eq:Lambda0}
\end{equation}
Here  $\lambda_{{\rm ab,}i}^{-2}(0)$ is the contribution of $i-$th band to $\lambda^{-2}_{\rm ab}(0)$, $L_{\rm c}$ is the $c-$axis lattice constant, ${\bf k}$ is a momentum vector, and $v_{\rm F}$ is the Fermi velocity. Integrations are made over the corresponding Fermi surface contours.

\begin{figure}[htb]
\includegraphics[width=1.0\linewidth]{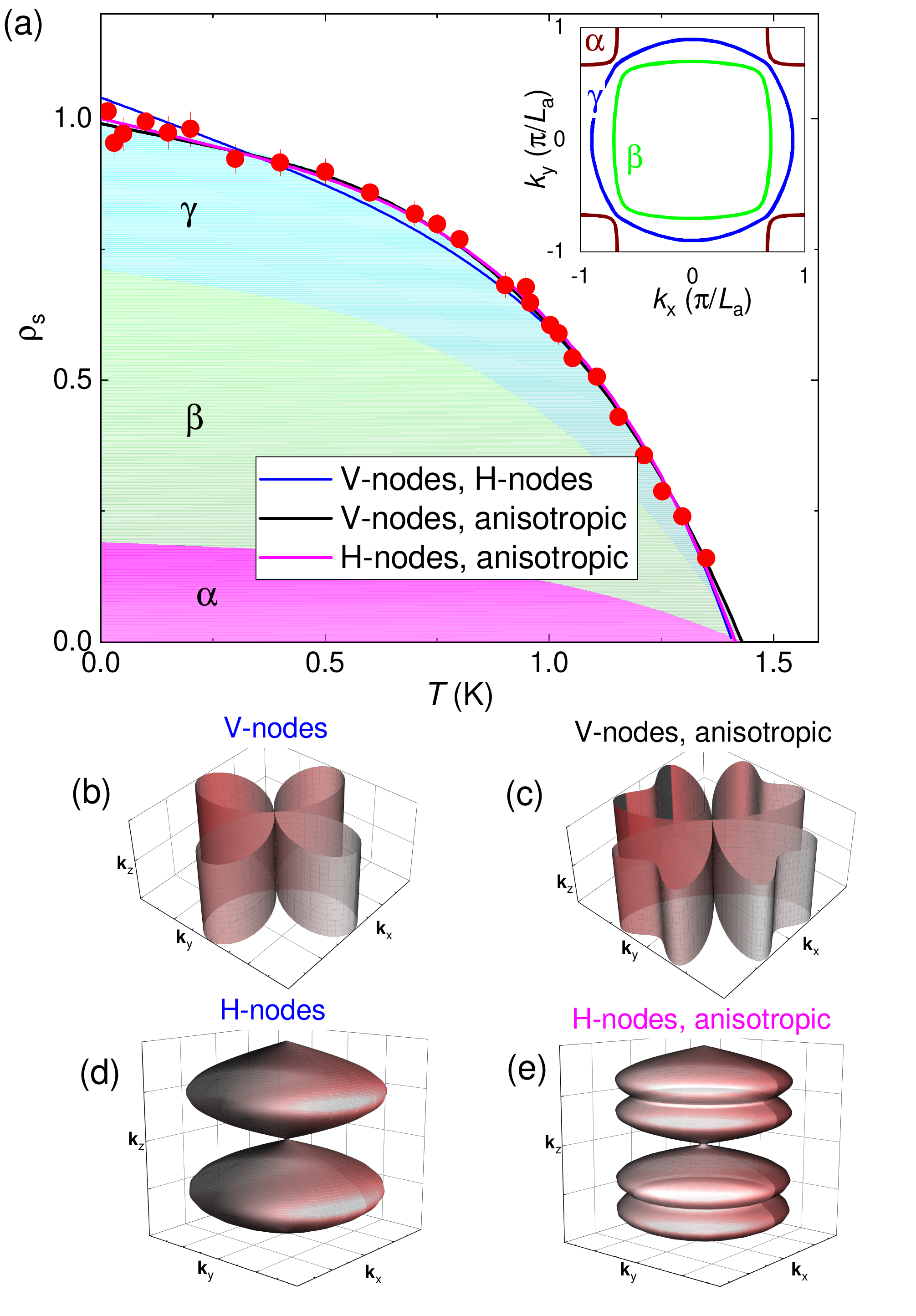}
\caption{(a)  The temperature evolution of the superfluid density $\rho_s(T)=\lambda_{\rm ab}^{-2}(T)/\lambda_{\rm ab}^{-2}(0)$ of Sr$_2$RuO$_4$. The solid lines are fits using different vertical (V) and horizontal (H) nodal gap structures (see text for details). The colored areas represent contributions of the $\alpha$, $\beta$, and $\gamma$  Fermi surface sheets. The inset shows the first Brillouin zone of Sr$_2$RuO$_4$ with the $\alpha$, $\beta$, and $\gamma$ Fermi surface sheets. $L_{\rm a}$ is the $a-$axis lattice constant. (b), (c), (d), and (e) different nodal gap structures fitted to the experimental $\rho_s(T$) data (see text for details).}
 \label{fig:superfluid-density}
\end{figure}

The Fermi surface of Sr$_2$RuO$_4$ consists of three nearly cylindrical sheets: $\alpha$, $\beta$, and $\gamma$ [the inset in Fig.~\ref{fig:superfluid-density}~(a)]. Precise ARPES measurements of the band structure and Fermi velocities were performed recently by Tamai {\it et al.}\cite{Tamai_PRX_2019} Using the band structure data from Ref.~\onlinecite{Tamai_PRX_2019} the weight of each Fermi surface sheet to $\lambda_{\rm ab}^{-2}(0)$ was found to be: $\omega_\alpha\simeq 0.19$, $\omega_\beta\simeq0.52$, and $\omega_\gamma\simeq 0.29$ [$\omega_i= \lambda_i^{-2}(0)/\lambda^{-2}(0)]$. The absolute value of $\lambda_{\rm ab}$ at $T=0$ was estimated as $\lambda_{\rm ab}(0)\simeq 130$~nm, in a good agreement with $\lambda_{\rm ab}(0)=124(3)$~nm obtained experimentally.

Figure~\ref{fig:superfluid-density}~(a) shows the temperature evolution of the superfluid density, $\rho_s(T)=\lambda_{\rm ab}^{-2}(T)/\lambda_{\rm ab}^{-2}(0)$. The data were reconstructed from $\lambda_{\rm ab}^{-2}(T)$'s  measured at $B_{\rm ap}=5$ and 20~mT (Fig.~\ref{fig:lambda_vs_T}).
In the clean limit, the temperature dependence of $\rho_{\rm s}(T)$ can be theoretically estimated by: \cite{Tinkham_75}
\begin{equation}
\rho_{\rm s}(T) =  1
+2\int_{\Delta(T,\varphi)}^{\infty}\frac{\partial
f(E,T)}{\partial E}D(E,T) dE~,
 \label{eq:superfluid}
\end{equation}
where $f(E,T)=[1+\exp(E/k_{\rm B}T)]^{-1}$ is  the Fermi function, $k_{\rm B}$ is the Boltzmann constant, and $D(E,T)$ is the density of states (DOS) in the superconducting state normalized by its value in the normal state. For an $s-$wave superconductor $D(E,T) = E/\sqrt{E^2-\Delta^2(T)}$, where $\Delta(T)$ is the magnitude of the superconducting gap at temperature $T$. A fully gapped superconductor has no DOS for $E<\Delta(T)$. As a consequence, the integral in Eq.~\ref{eq:superfluid} leads to an exponentially activated behaviour for the superfluid density at the lowest temperatures $\rho_{\rm s} (T) - \rho_{\rm s} (0)\propto e^{-\Delta(0)/(k_BT)}$. Conversely, for a nodal superconductor, the DOS is obtained by an average over the Fermi surface, $ \langle E/\sqrt{E^2-\Delta^2(T, \mathbf{k})}\rangle_{FS}$, where $\Delta(T, \mathbf{k})$ is temperature- and momentum-dependent, with $\Delta(T, \hat{\mathbf{k}}_n) = 0$ along the nodal directions. The presence of nodes allows for quasiparticle excitations at arbitrarily low temperatures and fundamentally changes the result of the integral in Eq. \ref{eq:superfluid}. In particular, for line nodes one finds $\rho_{\rm s}(T) - \rho_{\rm s}(0)  \propto  T$ and for point nodes  $\rho_{\rm s}(T) - \rho_{\rm s}(0)  \propto  T^2$, assuming a linearly vanishing gap at the nodes in both cases. \cite{Gross_86}
As the superfluid density is associated with a Fermi surface average, it is impossible to determine the position of nodes in momentum space solely from the information stemming from the $\rho_{\rm s}(T)$.  Nevertheless, the linear $T$-dependence of $\rho_{\rm s}(T)$ clearly suggests the presence of line nodes.

Fits of Eq.~\ref{eq:superfluid} to the experimental data were performed under the assumption of a quasi 2D (cylindrical) Fermi surface of Sr$_2$RuO$_4$.\cite{Mackenzie_RMP_2003} The superconducting gap function was assumed to depend on $T$ and spherical angles $\varphi$ and $\theta$ as: $\Delta(T,\varphi,\theta)=\Delta \ g(\varphi,\theta) \tanh\{1.82[1.018(T_c/T-1)]^{0.51}\} $ [$\Delta$ is the gap value at $T=0$ and $g(\varphi,\theta)$ is the angular dependence of the gap].\cite{Khasanov_PRL_La214_07} The presence of line nodes in the energy gap was accounted for by using following $g(\varphi,\theta)$ functions:  $g_{\rm V}(\varphi)=\cos(2\varphi)$ and $g_{\rm V,an}(\varphi)=a_{\rm V} \cos(2\varphi)+ (1-a_{\rm V})\cos(6\varphi)$ for vertical (V) nodes, and $g_{\rm H}(\theta)=\sin(\theta)$ and $g_{\rm H,an}(\theta)=a_{\rm H}|\sin(\theta)|+(1-a_{\rm H})|\sin(2\theta)|$ for horizontal (H) nodes, respectively. Here $a_{\rm V}$ and $a_{\rm H}$ are adjustable parameters. Note that $g_{\rm V}(\varphi)$ corresponds to the 'classical' gap of $d-$wave symmetry, while the 'anisotropic` $g_{\rm V,an}(\varphi)$ and $g_{\rm H,an}(\varphi)$ functions contain terms corresponding to the next angular harmonic.\cite{Mesot_PRL_1999}

The results of the fits are shown by solid lines in Fig.~\ref{fig:superfluid-density}~(a).\cite{comment} The corresponding $\Delta(\varphi,\theta)$ distributions are presented in panels (b), (c), (d), and (e). Note that the gap functions containing  only the lowest possible harmonic [$\cos(2\varphi)$ or $\sin(\theta)$] do not agree with the experimental $\rho_s(T)$ data, while both anisotropic functional forms describe the data equally well. The maximum gap values and the adjustable parameters are: $\Delta_{\rm V}^{\rm max}=\Delta_{\rm H}^{\rm max}=0.388$~meV; $\Delta^{\rm max}_{\rm V,an}=0.324$~meV, $a_{\rm V}= 1.47$; and $\Delta^{\rm max}_{\rm H,an}=0.348$~meV, $a_{\rm H,an}= 0.55$ for the above described gap models. Note that the maximum gap values obtained from fits ($\Delta^{\rm max}\simeq 0.32-0.39$~meV) stay in agreement with that measured directly by tunneling experiments ($\simeq 0.28-0.35$~meV, Refs.~\onlinecite{Sharma_PNAS_2020, Suderow_NJP_2009}).

From the results presented above, the following points should be highlighted:
(i)  The temperature dependence of the superfluid density is equally well described by both, vertical and horizontal, line nodes [Fig.~\ref{fig:superfluid-density}~(a)].
(ii) The Volovik effect (the inset in Fig.~\ref{fig:lambda_vs_T}) is active for a magnetic field component perpendicular to the Fermi velocity  $v_{\rm F}$.
% of quasiparticles around the nodes.
%, where the Doppler shift of a quasiparticle spectrum induces a finite DOS at the nodal region.
For $B_{\rm ap}\parallel c$ and for a cylindrical Fermi surface, $v_{\rm F}$s point radially away from the center of the cylinder, so the Doppler effect shifts the nodes around the Fermi surface in the azimuthal direction. For vertical line nodes, the Doppler shift induces a finite quasiparticle DOS at the nodes and the Volovik effect is expected. For horizontal line nodes the Doppler shift takes the nodal structure into itself and does not introduce a finite density of states at the nodes. A more realistic treatment with three Fermi surfaces for Sr$_2$RuO$_4$ could lead to an observable Volovik effect even for horizontal line nodes.\cite{Bang_PRL_2010, Wang_PRB_2011} The measurement of the penetration depth as a function of angles  $\varphi$ and $\theta$ may allow to probe the gap anisotropy based on the angular dependence of the Volovik effect.
(iii) Fits of Eq.~\ref{eq:superfluid} to $\rho_{\rm s}(T)$ assume the same gap magnitude for all three ($\alpha$, $\beta$, and $\gamma$) bands. If one allows the gap magnitudes to be different ({\it i.e.} $\Delta^{{\rm max,}\alpha}\neq\Delta^{{\rm max,}\beta}\neq\Delta^{{\rm max,}\gamma}$), and fit $\rho_{\rm s}(T)$ following the weight determined by the electronic band structure, with $\rho_s(T)=\sum_i \omega_i \rho_{s,i}(T)$ ($i=\alpha$, $\beta$, or $\gamma$), no qualitative difference with the single-gap fits are found. The maximum gap values remain the same within $\sim 10-20$\% accuracy.

%Comment of specific theoretical proposals
Recently, multiple theoretical proposals focusing on even-parity superconducting states have emerged after the report on spin-singlet superconductivity by nuclear magnetic resonance experiments.\cite{Pustogow2019, Chronister2021} These proposals fall into three categories: (i) accidentally degenerate order parameters, such as $s\pm i d$ \cite{Romer2019} or $d\pm i g$ \cite{Kivelson2020}; (ii) symmetry-protected two-component order parameters, such as chiral $d-$wave; \cite{Suh2020,Clepkens2021} (iii) single component order parameters with $d-$wave symmetry in the $B_{1g}$ channel.\cite{Willa2021}   Scenarios (i) and (iii) propose order parameters with vertical line nodes (or near nodes for the $s \pm i d$ scenario). Theoretical work based on spin-fluctuation mediated pairing,\cite{Romer2019} and calculations based on the random-phase approximation \cite{Wang2022} find order parameters with strong azimuthal angular dependence.  In contrast, scenario (ii) proposes order parameters with horizontal line nodes accompanied by vertical near nodes, \cite{Ramires2022} with unexpected angular dependence. All these scenarios are qualitatively compatible with the results reported here, in particular with the linear temperature dependence of $\lambda_{\rm ab}^{-2}$ at low temperatures, and gap anisotropies that cannot be captured by the lowest angular harmonic in a given symmetry channel.

%Conclusions
To conclude, we report on measurements of the in-plane magnetic penetration depth %($\lambda_{\rm ab}$)
in high-quality single crystals of Sr$_2$RuO$_4$ down to ultra-low temperature ($\simeq 0.015$~K) by means of the muon-spin rotation/relaxation.  The temperature evolution of $\lambda_{\rm ab}^{-2}$ represents pronounced linear behavior for temperatures $T\lesssim T_{\rm c}/2$, in agreement with the presence of line nodes or deep minima in the superconducting gap.  This picture is further confirmed by the reduction of $\lambda_{\rm ab}^{-2}(0)$ with increasing applied field, a manifestation of the Volovik effect.\cite{Volovik_JETP-L_1993, Kopnin_JETP-L_1996} The zero-field and zero-temperature value of $\lambda_{\rm ab}=124(3)$~nm
stays in agreement with $\lambda_{\rm ab}\simeq 130$~nm calculated based on the electronic structure measured by means of ARPES.\cite{Tamai_PRX_2019} The analysis of the temperature evolution of the superfluid density, reconstructed from the measured $\lambda^{-2}_{\rm ab}(T,B_{\rm ap})$ curves, reveals that the clean-limit behavior associated with the presence of a simple nodal superconducting gap described by the lowest possible angular harmonic does not account for the experimental data in the low temperature regime. To account for the low temperature behaviour, higher angular harmonics need to be introduced such that the low energy quasiparticle density of states is enhanced at low energies.

RK acknowledges helpful discussion with Clifford W. Hicks and support of Robert Scheuermann during $\mu$SR experiments. YM, VG, and NK are supported by JSPS Core-to-Core Program (No. JPJSCCA20170002). YM is supported by JSPS KAKENHI (No. JP22H01168), and NK is supported by JSPS KAKENHI  (Nos. JP18K04715, JP21H01033, and JP22K19093). AR acknowledges support from the Swiss National Science Foundation (SNSF) through an Ambizione Grant No.\ 186043. Z.G. acknowledges support from the Swiss National Science Foundation (SNSF) through SNSF Starting Grant (No. TMSGI2${\_}$211750).


\begin{thebibliography}{99}
\bibitem{comment} The fit of Eq.~\ref{eq:superfluid} to the data require the value $\rho(0)$ to be model dependent and to deviate from unity. This is the case for $g_{\rm V}(\varphi)=\cos(2\varphi)$ and $g_{\rm H}(\theta)=\sin(\theta)$ gap models.
\end{thebibliography}
\end{document}